\documentclass[twocolumn,
secnumarabic, amssymb, amsmath, nofootinbib,tightenlines,
showpacs,superscriptaddress,preprintnumbers, nobibnotes, aps,
prb]{revtex4}
\usepackage{amsfonts}
\usepackage{docs}%
\usepackage{bm}%
\usepackage[colorlinks=true,linkcolor=green,pagecolor=blue,filecolor=blue,menucolor=yellow,urlcolor=blue,citecolor=blue,anchorcolor=blue]{hyperref}%
\usepackage{graphicx}
\usepackage{dcolumn}
\usepackage{times}

\newcommand{\eg}{\textit{e.g. }}
\newcommand{\etal}{\emph{et al.}}
\def\i{\mathrm{i}}

\begin{document}

\title{Spin-triplet Supercurrent through Inhomogeneous Ferromagnetic Trilayers}

\author{Mohammad Alidoust}
\author{Jacob Linder}
\affiliation{Department of Physics, Norwegian University of Science
and Technology, N-7491 Trondheim, Norway}

\date{\today}

\begin{abstract}
Motivated by a recent experiment [J. W. A. Robinson, J. D. S. Witt
and M. G. Blamire, Science, \textbf{329}, 5987 (2010)], we here
study the possibility of establishing a long-range spin-triplet
supercurrent through an inhomogeneous ferromagnetic region
consisting of a Ho$\mid$Co$\mid$Ho trilayer sandwiched between two
conventional $s$-wave superconductors. We utilize a full numerical
solution in the diffusive regime of transport and study the behavior
of the supercurrent for various experimentally relevant
configurations of the ferromagnetic trilayer. We obtain
qualitatively very good agreement with experimental data regarding
the behavior of the supercurrent as a function of the width of the
Co-layer, $L_\text{Co}$. Moreover,
we find a synthesis of 0-$\pi$ oscillations with superimposed rapid
oscillations when varying the width of the Ho-layer which pertain specifically to the spiral magnetization texture in Ho. We are not able
to reproduce the anomalous peaks in the supercurrent observed experimentally in this
regime, but note that the results obtained are quite sensitive to
the exact magnetization profile in the Ho-layers, which could be the reason
for the discrepancy between our model and the experimental reported
data for this particular aspect. We also investigate the
supercurrent in a system where the intrinsically inhomogeneous Ho
ferromagnets are replaced with domain-wall ferromagnets, and find
similar behavior as in the Ho$\mid$Co$\mid$Ho case. Furthermore, we
propose a novel type of magnetic Josephson junction including only a
domain-wall ferromagnet and a homogeneous ferromagnetic layer, which in addition to
simplicity regarding the magnetization profile also offers a tunable long-range
spin-triplet supercurrent. Finally, we discuss some experimental aspects of our findings.
\end{abstract}


\maketitle

\section{Introduction}
Due to the rich physics from a fundamental viewpoint and the
possibility of birthing practical applications, configurations
containing superconductors and ferromagnets have attracted much
attention
theoretically\cite{khaire1,khaire2,khaire3,kontos1,kontos2,klapwijk,wang}
and
experimentally\cite{buzdin1,bergeret1,golubov1,tanaka1,tanaka2,linder1,eschrig1,buzdin2}
over the last years (see for instance Ref. \onlinecite{bergeret1}
for a comprehensive review and reference-list).
One of the most intriguing phenomena in the context of this
interplay is the generation of a long-ranged spin-triplet
supercurrent flowing through a Josephson junction with magnetic
elements. The main criterion for generation of such a current is
that some form of magnetic inhomogeneity must be present in the
junction, as first shown by Bergeret \textit{et
al.}\cite{bergeret2,bergeret3,bergeret4} and Volkov \textit{et al.}
\cite{volkov1,volkov2,volkov3}. In the diffusive limit of transport,
being often the experimentally most relevant regime, such a
non-uniform magnetization can  induce exotic long-range
superconducting correlations which are odd under time-reversal: the
so-called odd-frequency superconducting state. The resulting
triplet-supercurrent decays over same length scale as in a
superconductor$\mid$normal metal$\mid$superconductor
(S$\mid$N$\mid$S) junction, but now additionally activates the
spin-degree of freedom in the transport of Cooper pairs. A
long-range triplet-supercurrent was predicted to occur in a setup
consisting of three non-collinear homogeneous magnetic layers
\cite{buzdin2}, but disappears in the scenario of only two
homogeneous non-collinear magnetic layers, as discussed in Refs.
\onlinecite{blanter,crouzy,robinson1,sperstad2}. Over the last
couple of years, induction of triplet-correlations in layered
heterostructures with ferromagnets (F) and superconductors (S) in
the clean limit has been studied by using different formalisms and
configurations in Refs. \onlinecite{halterman1,halterman2,radovic}.
On the experimental side, Keizer \etal\cite{keizer} observed a
long-range supercurrent through half-metallic CrO$_2$, whereas very
recent work also reports observation of a long-range supercurrent
through an inhomogeneous magnetic layer \cite{khaire1, sprungmann}.
In particular, Robinson \etal\cite{robinson1} investigated the
appearance of a spin-triplet supercurrent flowing through a magnetic
Ho$\mid$Co$\mid$Ho trilayer. Due to the intrinsic magnetic
inhomogeneity in Ho, featuring a spiral magnetization texture, it
was found that a strong spin-triplet supercurrent was established
through the trilayer connecting two $s$-wave superconducting leads.
In this paper, motivated by the very recent experimental in Ref.
\onlinecite{robinson1}, we utilize a full numerical solution of the
quasiclassical Green's function in the diffusive regime and study
theoretically spin-triplet condensation in the critical charge
current flowing through a Ho$\mid$Co$\mid$Ho magnetic trilayer. Due
to our numerical approach, we have access to the full-proximity
effect regime and complicated magnetization textures in the
trilayer. This allows us to also study the influence of domain-walls
in the ferromagnet on the behavior of the long-range supercurrent.

The main results in Ref. \onlinecite{robinson1} due to Robinson
\etal\;were (i) a slow decay of the supercurrent as a function of
the Co-layer thickness and (ii) anomalous peaks arising in the
characteristic voltage of the junction as a function of the Ho-layer
thickness. Using the computational machinery described above, we
obtain qualitatively very good agreement with the experimental data
pertaining to (i). However, we are not able to reproduce the
anomalous peaks observed for the supercurrent pertaining to (ii).
Instead, we find a synthesis of 0-$\pi$ oscillations with
superimposed rapid oscillations which pertain specifically to the
spiral magnetization texture in Ho. However, we also show how the
exact behavior of the supercurrent vs. the width of the
inhomogeneous magnetic layer is rather sensitive to the exact
magnetization pattern. This suggests that the trilayer magnetization texture
realized in the experiment by Robinson \etal\cite{robinson1} might
differ somewhat from our model.
Motivated by the above mentioned reason, we also investigate how the critical current behaves when we replace the Ho layers in the
trilayer junctions with domain-walls. Finally, we propose a
novel type of inhomogeneous ferromagnetic Josephson
junction with a simpler magnetization profile compared to previous proposals, including only a domain-wall and homogeneous ferromagnet, and demonstrate the possibility to tune the long-ranged spin-triplet
supercurrent flowing through the junction.
\par
This work is organized as follow: In Sec. \ref{theory}, we present
the main ingredients of the theory which is used throughout the
paper, i.e. a quasiclassical Green's function method in the
diffusive limit studied by means of the Usadel equation and
supplemented with proper boundary conditions. In Sec. \ref{HCH}, we
investigate the triplet-supercurrent flowing through an
Ho$\mid$Co$\mid$Ho ferromagnetic trilayer as a function of both the
Ho- and Co-layer thickness including two different magnetization
textures in the Ho layers. In Sec. \ref{DCD}, we investigate how the
behavior of the supercurrent is altered when the Ho regions are
replaced with domain-wall ferromagnets, which also feature an
intrinsical inhomogeneous magnetization texture. We also propose
a novel type of ferromagnetic Josephson junctions to investigate the possibility of tuning the long-range contribution to the supercurrent via an external field.
Finally, we summarize
and give concluding remarks in Sec. \ref{summary}.

\begin{figure}[t!]
\includegraphics[width=8.5cm,height=8cm]{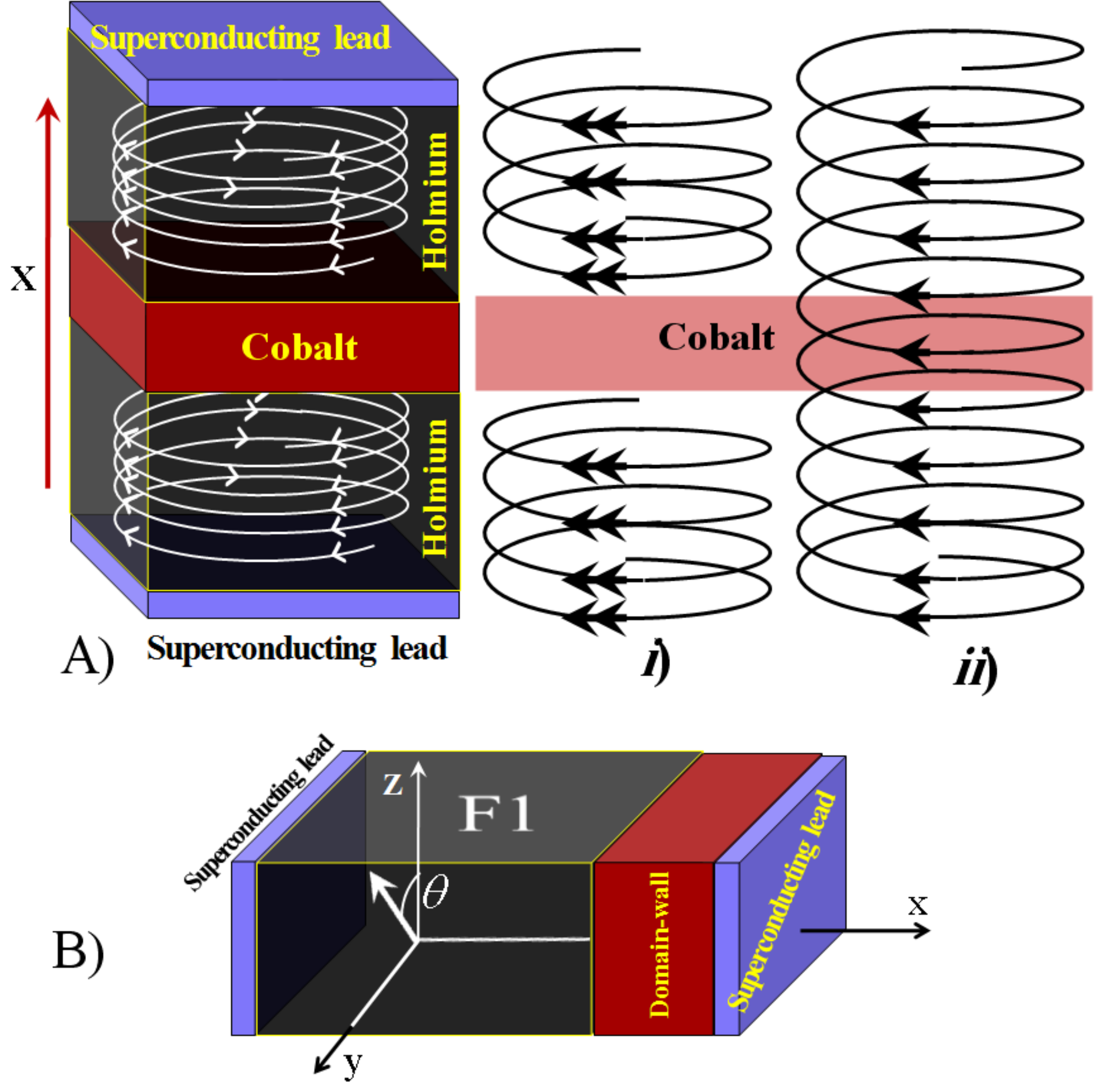}
\caption{\label{fig:model} (Color online) A): The schematic setup of
a ferromagnetic trilayer of Ho$\mid$Co$\mid$Ho. The spiral curves
show the trajectory of the magnetization vector in the Ho layer
which rotates along a conical profile in the $\hat{x}$-direction.
\textit{i}) shows the configuration in which the magnetization
patterns of the two Ho layers are completely identical, whereas in \textit{ii}) the
magnetization patterns follow a continuous spiral magnetization in
the two Ho layers. B): Schematic model
of the experimental setup of our proposed ferromagnetic Josephson
junction including a domain-wall and homogeneous ferromagnetic layer
which enables a controllable triplet-supercurrent. The angle $\theta$
represents the orientation of the homogeneous magnetization in the F1 layer
with respect to the $\hat{z}$-direction. }
\end{figure}

\section{Theoretical approach}\label{theory}

For studying various characteristics of different media, the Green's
function method is a fundamental approach utilized in many areas of
condensed-matter physics \cite{mahan,pitaevski}. In the context of
non-equilbrium transport through different media one should start
from Dyson's equation of motion and calculate the Keldysh Green's
function\cite{venkat}. In equilibrium situations, the Keldysh block
of Green's function can be obtained from the Advanced and Retarded
blocks of the Green's function. Inside superconducting regions,
Dyson's equation of motion equation transforms to Gorkov's equations
which in turn can be reduced to Eilenberger's equation within a
quasiclassical approximation, where the Fermi wavelength is much
smaller than all other length scales. The Eilenberger equation
reads: \cite{eilenberger}
\begin{equation}
[E\hat{\tau}_3+\hat{\Delta},\hat{G}]+\i \boldsymbol{v}_F\cdot\nabla
\hat{G}-[\hat{\Sigma},\hat{G}]=0,
\end{equation}
in which $\boldsymbol{v}_F$ is vector Fermi velocity of the 
quasi-particles in the superconducting region and $\hat{\Sigma}$ is a 
self-energy term related to \eg elastic and spin-flip
scattering centers. In the contrast of real part of $\hat{\Sigma}$
which is an oscillatory function of energy, the imaginary part of
the term is dissipative and decays when increasing the energy. Here,
$\hat{\Delta}$ and $\hat{\tau}_3$ are defined as:
\begin{equation}\label{eilenberger}
\nonumber\hat{\Delta}=\left(\begin{array}{cc}
\underline{0} & \underline{\tilde{\Delta}}  \\
\underline{\tilde{\Delta}}^* & \underline{0} \\
\end{array}\right),\:\tilde{\Delta}=\left(\begin{array}{cc}
0 & \Delta \\
-\Delta & 0
\end{array}\right),\:\hat{\tau_3}=\left(\begin{array}{cc}
\underline{1} & \underline{0} \\
\underline{0} & -\underline{1}
\end{array}\right).
\end{equation}
Hwhile $\underline{\Box}$ and $\hat{\Box}$ stand for $2\times 2$ and
$4\times 4$ matrix quantities.

In experimental situations, the diffusive regime of transport is
often reached as very clean (ballistic) samples may be hard to
fabricate. In the diffusive limit, impurities in specimen are very
strong and consequently the self-energy term $\hat{\Sigma}$ in the
Eq. (\ref{eilenberger}) dominates. The impurities will lead to
randomization of the quasiparticle trajectories, scattering them all
over $k$-space. By expanding the Green's function with respect to
$s$- and $p$-wave spherical harmonics and performing an angular
averaging process, one arrives at the Usadel-equation\cite{usadel}:
\begin{align}\label{eq:usadel}
\nabla(\hat{G}\nabla\hat{G}) + \frac{\i}{E_{th}}[E\hat{\rho}_3 +
\text{diag}
[\boldsymbol{h}\cdot\underline{\boldsymbol{\sigma}},(\boldsymbol{h}\cdot
\underline{\boldsymbol{\sigma}})^{T}], \hat{G}]=0,
\end{align}
where $E_{th}=D/{d_F}^{2}$ is Thouless energy in which $D$ is
diffusive constant and $d_F$ is length of ferromagnetic layer, $h$
is exchange field of ferromagnetic region and $\hat{\rho_3}$,
$\underline{\boldsymbol{\sigma}}$ are Pauli matrixes which are
available in Appendix \ref{pauli}.

Throughout the paper, we shall assume that the ferromagnetic layer
has been sandwiched between two conventional $s$-wave
superconducting leads whose interfaces are located at $x=-d_F/2$ and
$d_F/2$. Due to the isotropic superconducting order parameter and
the impurity scattering, we may capture all the essential physics by
considering an effective one dimensional system, thus
$\nabla\equiv\partial/\partial_x\equiv\partial_x$ in Eq.
(\ref{eq:usadel}). For investigating the charge-current flowing
through the system, we employ the following boundary conditions at
the two contact regions with the superconducting reservoirs:

\begin{align}\label{bc}
 \left\{\begin{array}{cc}
2\zeta\hat{G}\partial_x\hat{G}=[\hat{G}_{\text{BCS}}(\phi),\hat{G}]   \qquad x=-\frac{d_F}{2}\\
2\zeta\hat{G}\partial_x\hat{G}=[-\hat{G}_{\text{BCS}}(-\phi),\hat{G}]
\qquad x=\frac{d_F}{2}
\end{array}
    \right.
\end{align}
where $\hat{G}_{\text{BCS}}$ is the bulk Green's function in the
superconductors and $\zeta$ is defined as ratio between the
resistance of the barrier region ($R_B$) and the resistance in the
ferromagnetic film ($R_F$). We disregard here the influence of spin-dependent interfacial phase-shifts occuring at the interfaces since their effect is unimportant in the present context of an intrinsically inhomogeneous magnetization structure (including them would introduce slight shifts to the 0-$\pi$ transition points). \cite{huertas-hernando_prl_02, cottet_prb_05}.

For solving the Usadel equation and implementing boundary conditions
numerically, it is convenient to parameterize the Green's function.
There are two standard parameterizations approaches; $\theta$- and
Ricatti-parameterizations, and we will here employ the latter.
The parameterized Green's function then reads as follows:
\begin{align}\label{eq:gricatti}
\hat{G}=\left(
  \begin{array}{cc}
    \underline{N}(\underline{1}-\underline{\gamma}\underline{\widetilde{\gamma}}) & 2\underline{N}\underline{\gamma} \\
    2\underline{\widetilde{N}}\underline{\widetilde{\gamma}} & \underline{\widetilde{N}}(-\underline{1}+\underline{\widetilde{\gamma}}\underline{\gamma}) \\
  \end{array}
\right).
\end{align}
By imposing a normalization condition for the Green's function,
namely $\hat{G}^2=\hat{1}$, $\underline{N}$ and
$\underline{\widetilde{N}}$ are obtained as
\begin{align}\label{normalization_con}
\underline{N}=\frac{1}{\underline{1}+\underline{\gamma}\underline{\widetilde{\gamma}}}\qquad\underline{\widetilde{N}}=\frac{1}{\underline{1}+\underline{\widetilde{\gamma}}\underline{\gamma}}.
\end{align}
Within the Ricatti-parametrization scheme, the components of the
bulk superconductor Green's function are:
\begin{align}
\gamma_{\text{BCS}}(\phi) &= i\underline{\tau_2}s/
(1+c)e^{i\phi/2},\notag\\
\widetilde{\gamma}_\text{BCS}(\phi) &=
\gamma_{\text{BCS}}(\phi)e^{-i\phi},
\end{align}
where $\phi$ is superconducting phase difference between the two
$s$-wave superconducting leads and $s$, $c$ is defined as
$\sinh\vartheta$ and $\cosh\vartheta$, respectively, in which
$\vartheta=\text{atanh}(\Delta(T)/E)$. We use standard BCS
temperature dependent of superconducting gap in our calculations and
$\Delta_0=\Delta(0)$ stands for superconducting gap in the absolut
zero. Throughout the paper we normalize all energies with respect to
superconducting gap at the zero temperature ($\Delta_0$) and all
lengths with respect to ferromagnetic layer length. We use units so
that $\hbar=k_B=1$.

\begin{figure}[b!]
\includegraphics[width=8cm,height=6.5cm]{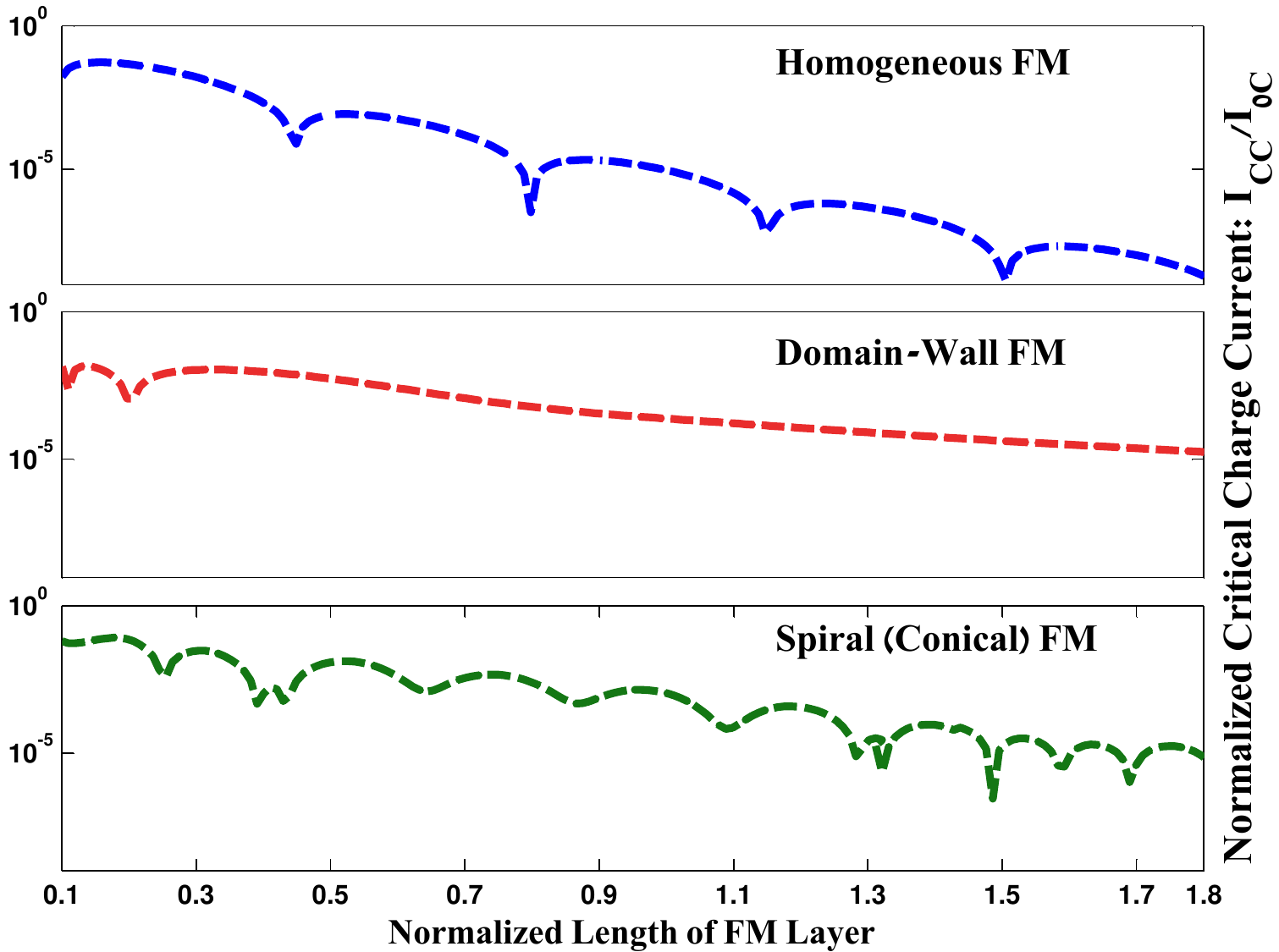}
\caption{\label{fig:pureFM} (Color online) The normalized critical
charge current for three types of magnetization textures. Top,
middle and bottom frames show critical supercurrent through
sandwiched uniform, domain-wall and spiral ferromagnetic layers,
respectively.}
\end{figure}
For investigating the electronic transport properties of all
configurations one needs to obtain the Keldysh block of the Green's
function. Under equilibrium conditions, the Keldysh component can be
obtained from the Retarded and Advanced blocks as
$\hat{G}^{K}=(\hat{G}^{R}-\hat{G}^{A})\tanh(E/2k_BT)$ and
$\hat{G}^{A}=-(\hat{\rho}_3\hat{G}^{R}\hat{\rho}_3)^{\dag}$. The
charge-current is then obtained via:
\begin{eqnarray}
\frac{I_{C}}{I_{0C}}= \left|\int_{-\infty}^{+\infty}dE\;
\text{Tr}\left\{\hat{\rho}_3\left(\hat{G}\partial_x\hat{G}\right)^{K}\right\}\right|
\end{eqnarray}
in which $I_{0C}=N_0eD/16d_F$, $N_0$ is the normal density of states
per spin.
Above, $\left(\hat{G}\partial_x\hat{G}\right)^{K}$ denotes the
Keldysh component of the $\hat{G}\partial_x\hat{G}$ matrix. We now proceed to study the transport
properties of several experimentally accessible
configurations, in particular pertaining the experiment by Robinson \etal\cite{robinson1}. We underline once more that our approach grants us access to the full proximity regime in addition to complicated magnetization textures which cannot be treated analytically.

\section{Spin-triplet supercurrent through a ferromagnetic Ho$\mid$Co$\mid$Ho trilayer}\label{HCH}

In this section, we present main results of the paper: a theoretical
investigation of a ferromagnetic Ho$\mid$Co$\mid$Ho trilayer
sandwiched between two $s$-wave superconductors, as recently
experimentally studied in Ref. \onlinecite{robinson1}. The
magnetization structure of isolated Ho is experimentally known
\cite{sosnin} and depicted in  portion A) of Fig. \ref{fig:model}.
Analytically, the instantaneous direction of the local magnetization
can thus be written as:
\begin{eqnarray}
 \boldsymbol{h}= h\;(\cos\alpha\;\hat{x}+\Xi\;\sin\alpha)\\
\nonumber\Xi=\left\{\sin(Qx)\;\hat{y}+\cos(Qx)\;\hat{z}\right\},
\end{eqnarray}

where $Q=2\pi/\lambda$ and $\lambda$ is the spiral length in Ho. In
Ref. \onlinecite{robinson1}, this was estimated to $\lambda\simeq
3.4$ nm. The apex angle is denoted by $\alpha$ and equals
$4\pi/9$.\cite{sosnin,robinson1} For both Ho and Co ferromagnetic
layers, the strength of the exchange field is larger than alloys and
compounds of Pd and Ni. For Pd$_x$Ni$_{1-x}$ alloys, weak strengths
of the exchange field in the range of $h/\Delta_0\approx 5-10$ are
accessible in the experiments. This should be contrasted with
intrinsically ferromagnetic materials whose exchange field strengths
are often very large (100 meV -- 1 eV). We set superconducting
coherent length as $\xi_S=15$ nm which is accessible for example
in Nb. For more stability in the numerical code, we add an imaginary
part to the quasi-particle energies equal to $\delta/\Delta_0=5\times
10^{-2}$ which modeling inelastic scattering in the
specimen. To simulate numerically the ferromagnetic trilayer
sandwiched between the two $s$-wave superconducting leads, we should
first note the parameter regime in which the quasi-classical Green's
function method is valid. Due to the requirement of the Fermi energy
being much larger than all energy scales, we set $h/\Delta_0=80$ to
model a strong exchange field still within the regime of validity.
Throughout our calculations, we fix the temperature at $T/T_c=0.2$
and also set the ratio of barrier and ferromagnetic resistances to
$\zeta=R_B/R_F=5$. For the value of the exchange field considered
here, it can be shown that the spin-dependent phase-shifts occurring
at the interface may be neglected.

To begin with, we mention briefly the qualitative difference between
having a homogeneous and inhomogeneous magnetization in the magnetic
layer. In Fig. \ref{fig:pureFM}, we show the critical charge current
behavior vs. the thickness of ferromagnetic layer for three
scenarios: a homogeneous exchange field, a domain-wall ferromagnet,
and finally a spiral (conical) magnetization texture. As seen, the
fundamental difference between the homogeneous case in the upper
panel of Fig. \ref{fig:pureFM} and the two inhomogeneous scenarios
is that the current becomes long-ranged in the latter cases, i.e. the
critical supercurrent decays on a much larger length-scale compared to the homogeneous case. This is due to the generation of a long-range triplet
supercurrent which is sustained by the inhomogeneous field
\cite{bergeret1,volkov1, eschrig1, halterman1, radovic}. Moreover,
the spiral magnetization pattern gives rise to a rapid oscillation
pattern superimposed on the 0-$\pi$ transitions, as was shown in
Refs. \onlinecite{alidoust, annett}.

\begin{figure}[t!]
\includegraphics[width=8.0cm,height=7.3cm]{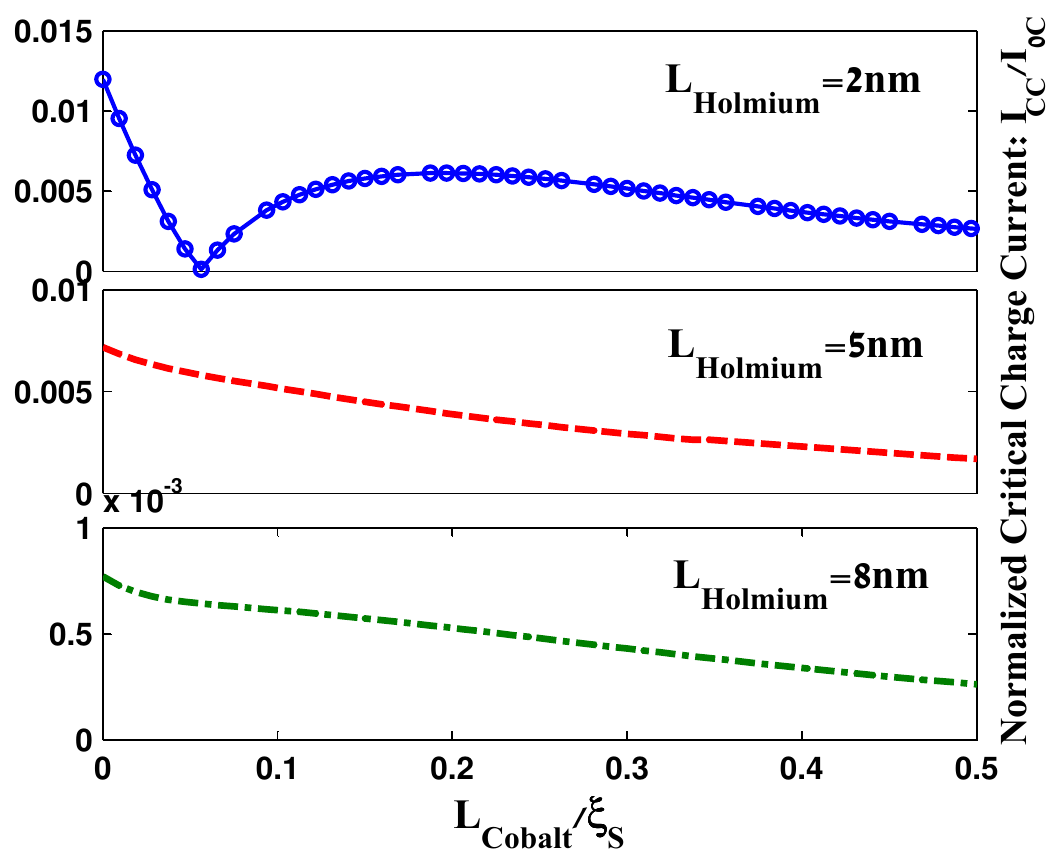}
\caption{\label{fig:J_vsLCo-Ho-Z} (Color online) The normalized
critical supercurrent through a magnetic Ho$\mid$Co$\mid$Ho trilayer
vs. the length of the Co layer for three values of Ho-layer lengths,
$L_{\text{Ho}}$=2 nm, 5 nm and 8 nm. }
\end{figure}

\begin{figure}
\includegraphics[width=8.0cm,height=7.5cm]{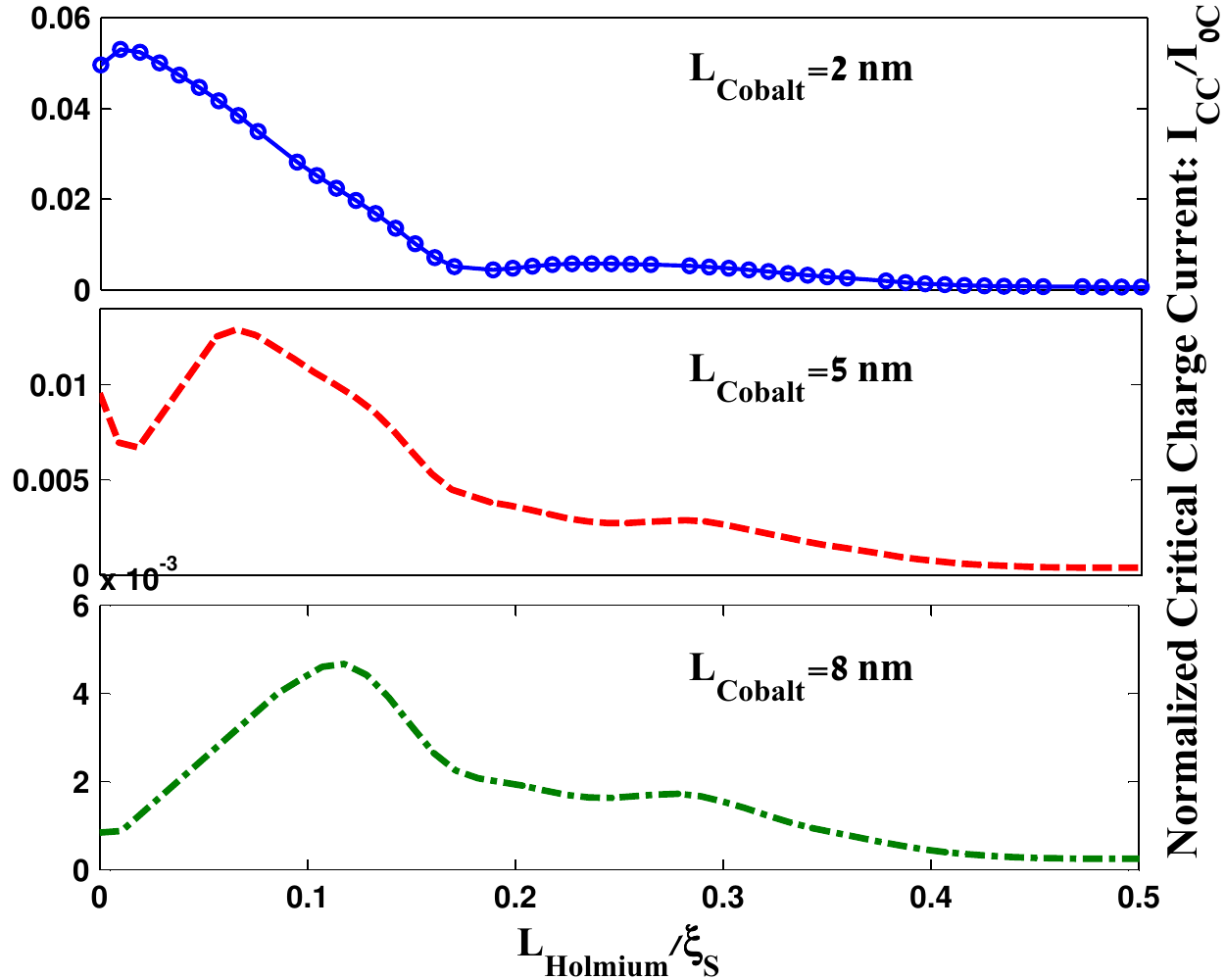}
\caption{\label{fig:J_vsLHo-Co-Z} (Color online) The normalized
critical supercurrent through a ferromagnetic trilayer of
Ho$\mid$Co$\mid$Ho vs. the length of the Ho layer for three values
of Co-layer lengths, $L_{\text{Co}}$=2 nm, 5 nm and 8 nm. }
\end{figure}

Now we model the ferromagnetic Ho$\mid$Co$\mid$Ho trilayer by:
\begin{eqnarray}\label{eq:mag-part-domain}
\boldsymbol{h} =
                                   \left\{
                                     \begin{array}{ccc}
                                        h\;(\cos\alpha\;\hat{x}+\Xi\;\sin\alpha)\;  & x<-\frac{L_{\text{Co}}}{2} \\
                                       h\;\hat{z}  & -\frac{L_{\text{Co}}}{2}<x<\frac{L_{\text{Co}}}{2}\;, \\
                                        h\;(\cos\alpha\;\hat{x}+\Xi\;\sin\alpha)\;  & x>\frac{L_{\text{Co}}}{2} \\
                                     \end{array}
                                   \right.
\end{eqnarray}
where we assume that the middle of the Co-layer is located at $x=0$. In Fig.
\ref{fig:J_vsLCo-Ho-Z}, we show the variations of the critical
supercurrent when the Co-layer length is varying [see portion A) of
Fig. \ref{fig:model} for the structure under consideration]. Results
are provided for three distinct values of the Ho-layers length. We
first assume that the Ho-layers have identical spiral magnetization
patterns. As we shall see later, the results for the
critical current are sensitive to the exact magnetization texture
in the Ho-layers. The
magnetization of the Co-layer is taken to be along the
$\hat{z}$-axis, \textit{i.e.} parallel with the contact-interfaces,
as should be reasonable for a thin-film structure. As seen in Fig.
\ref{fig:J_vsLCo-Ho-Z}, for thin Ho-layers, the critical
supercurrent displays only one 0-$\pi$ transition point. For larger
values of $L_\text{Ho}$, the supercurrent decays monotonically
similar to flowing critical supercurrent through an S$\mid$N$\mid$S
junction where only spin-singlet condensation contributes to the 
supercurrent. This is
consistent with the experimental observation by Robinson \etal
\cite{robinson1}. For thick Ho-layers, the charge supercurrent thus
behaves as if a normal layer has been sandwiched between two
$s-$wave superconducting. The long decay length of the supercurrent,
comparable to a normal Josephson junction, is evidence of precisely
a long-ranged spin-triplet supercurrent flowing through the
ferromagnetic Ho$\mid$Co$\mid$Ho trilayer.

\begin{figure}
\includegraphics[width=8.0cm,height=6.8cm]{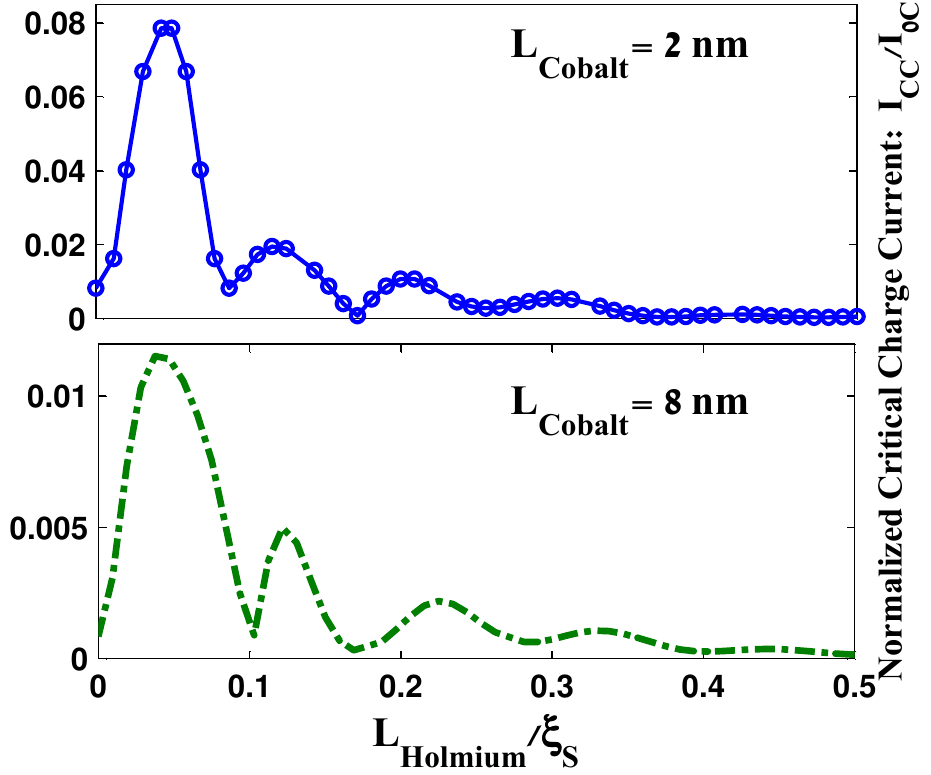}
\caption{\label{fig:M_vsLHo-Co-Z} (Color online) The normalized
critical supercurrent through a trilayer of Ho$\mid$Co$\mid$Ho vs.
the length of the Ho layer for two values of Co-layer lengths,
$L_{\text{Co}}$=2 nm and 8 nm. In this case, we assume that
the magnetization vector in Ho layers follow a continuous spiral
pattern without being interrupted by the Co layer.}
\end{figure}

Next, we investigate how the current behaves upon altering the
Ho-layer thickness. We show results for three distinct values of
$L_\text{Co}$ in Fig. \ref{fig:J_vsLHo-Co-Z}. The Co-layer exchange
field is as before assumed to be oriented parallell to the interface
regions. As seen, the current now decays in a non-monotonic fashion.
Interestingly, Robinson \etal \cite{robinson1} observed a set of
anomalous sharp peaks in the current when $L_\text{Ho}$ is
increasing. Although we are not able to reproduce such sharp peaks
within this quasiclassical treatment, we confirm the non-monotonic dependence of the critical current observed by Ref. \onlinecite{robinson1}. In order to investigate further if the specific
magnetization profile is crucial with regard to the appearance of
the anomalous behavior observed experimentally, we investigate a
slightly different magnetization texture model in the Ho-layers.
Whereas the magnetization pattern previously was assumed to be
identical in both layers, we show in Fig. \ref{fig:M_vsLHo-Co-Z}
results for the case when the magnetization pattern in the right Ho
layer couples continuously to the left layer as if the Co-layer has
no influence on it [see part \textit{ii}) of portion A) of Fig.
\ref{fig:model}]. Unlike the first case, several minima now appear
in the critical supercurrent, out of which two are 0-$\pi$
transition points while the others are local minima. \textit{This
demonstrates that the exact behavior of the supercurrent is
sensitive to the specific magnetization profile in the inhomogeneous
magnetic layers.} Motivated by this finding, we explore in the next
section how the results change when the Ho-layers are replaced by
domain-wall ferromagnets.


\begin{figure}
\includegraphics[width=8.0cm,height=7cm]{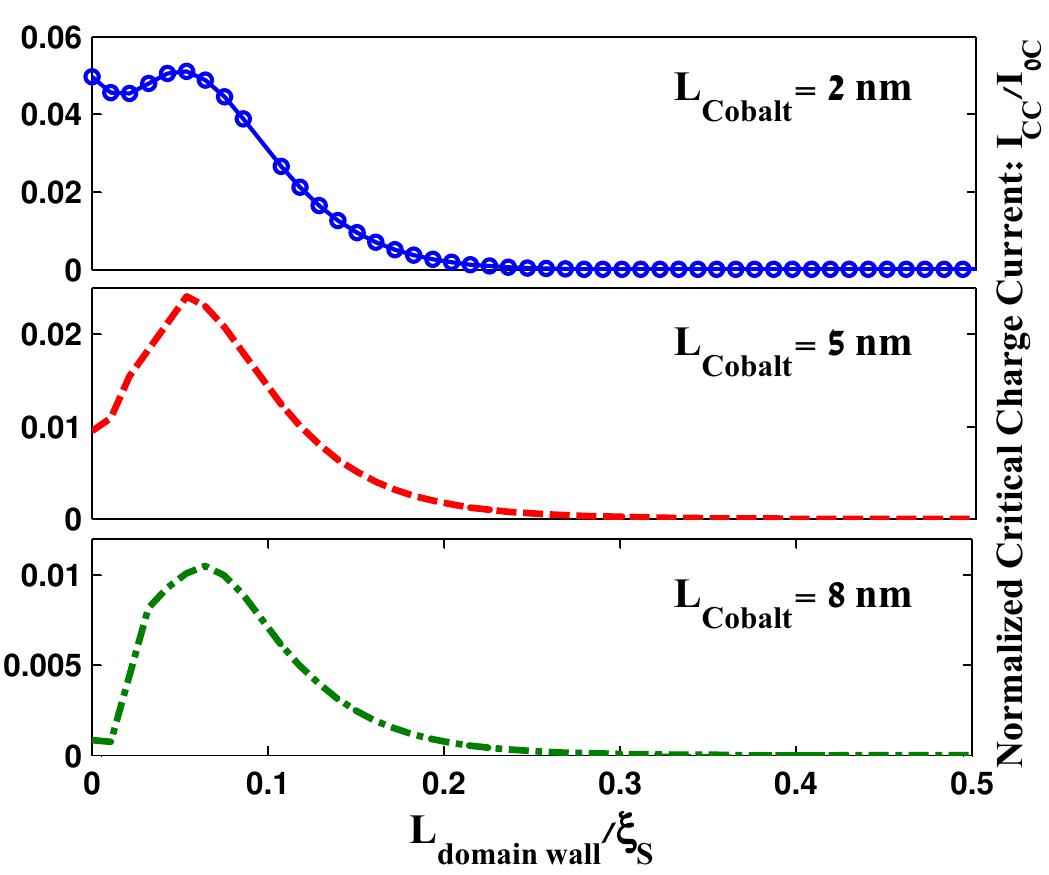}
\caption{\label{fig:Bloch-Bloch-vsLdomain-Z} (Color online) The
normalized critical supercurrent through sandwiched
Bloch-wall$\mid$Co$\mid$Bloch-wall trilayer vs length of Bloch-wall layers for
three different values of Co layer lengths, $L_{\text{Co}}$=2 nm, 5
nm and 8 nm. Here magnetization direction of Co layer is along the
$\hat{z}$-axes. }
\end{figure}

\section{Spin-triplet supercurrent through a ferromagnetic structure with domain-walls}\label{DCD}

We keep the same parameters as used in the previous section, but now
replace the spiral magnetization patterns in portion A) of Fig.
\ref{fig:model} with domain-wall ferromagnets. Both a Neel and
Bloch-wall configuration have been investigated by us numerically,
and were found to give virtually identical results for the
charge-current transport. Thus, we here present results only for the
Bloch domain-wall case. Fig. \ref{fig:Bloch-Bloch-vsLdomain-Z} shows
the critical charge-current through a ferromagnetic trilayer with
the following magnetization structure:

\begin{eqnarray}\label{eq:mag-part-domain}
\boldsymbol{h} =
                                   \left\{
                                     \begin{array}{ccc}
                                       h(\cos\beta\;\hat{y}+\sin\beta\;\hat{z})  & x<-\frac{L_{\text{Co}}}{2} \\
                                       h\;\hat{z}  & -\frac{L_{\text{Co}}}{2}<x<\frac{L_{\text{Co}}}{2} \\
                                       h(\cos\beta\;\hat{y}+\sin\beta\;\hat{z})  & x>\frac{L_{\text{Co}}}{2} \\
                                     \end{array}
                                   \right.
\end{eqnarray}
where $d_W$ is width of
domain-wall and:
\begin{align}
\beta=-2\text{atanh}\Big(\frac{x-L_{d}/2}{d_W}\Big).
\end{align}
We set $d_{W}$=$L_{d}/2$ throughout our computations, and provide
results for the critical supercurrent biased through the trilayer vs. the 
domain-wall$-$layer length $L_d$ for three distinct values of the Co-layer
lengths in Fig. \ref{fig:Bloch-Bloch-vsLdomain-Z}. For
$L_{\text{Co}}=2\; \text{nm}$ the critical supercurrent features two
0-$\pi$ transition points and these transition points disappear when
$L_\text{Co}$ increases. This finding coincides with the
Ho$\mid$Co$\mid$Ho trilayer. As can be seen from Fig.
\ref{fig:Bloch-Bloch-vsLdomain-Z}, the variation of critical
supercurrent vs. the length of domain-wall$-$layer shows a
non-monotonic behavior, as in Fig. \ref{fig:J_vsLHo-Co-Z}. In effect,
domain-wall ferromagnets can serve a similar purpose as Ho with
regard to the generation of a long-range current.

Due to the somewhat complicated magnetization texture in the trilayer structure considered in Ref. \onlinecite{robinson1}, it is tempting to consider if it is possible to
simplify the structure of the magnetic layers and still obtain a setup where the triplet
supercurrent can be experimentally controlled. In Ref.
\onlinecite{buzdin2}, a magnetic trilayer consisting of homogeneous,
misaligned ferromagnets was proposed as a setup where the long-range
current could be controlled by varying the angle of misalignment.
Experimentally, it would nevertheless be highly challenging to exert
individual control over the local magnetization field in each layer
via application of external fields. We here propose another type of
heterostructure which might be more beneficial in this regard. As is
seen in part B) of Fig. \ref{fig:model}, we consider a
superconductor$\mid$domain-wall$\mid$ferromagnet$\mid$superconductor
junction where the coercive field of the homogeneous ferromagnet is
sufficiently low to allow tuning of its magnetization via an
external field without altering the domain-wall ferromagnet.
In this
way, the domain-wall ferromagnet serves as a source for the
long-range triplet current, while the orientation of the homogeneous
ferromagnet can tune this contribution. We define the magnetic field
orientation angle of the F1 layer as $\theta$ (see Fig.
\ref{fig:model}, portion B)), and thus the magnetization profile of the
ferromagnetic layer reads:

\begin{eqnarray}\label{eq:mag-part-domain}
\boldsymbol{h} =
                                   \left\{
                                     \begin{array}{ccc}
                                         h_1(\cos\theta\;\hat{z}+\sin\theta\;\hat{y})& x<L_{\text{F1}} \\
                                       h_2(\cos\beta\;\hat{y}+\sin\beta\;\hat{z})  & x>L_{\text{F1}} \\
                                                                            \end{array}
                                   \right..
\end{eqnarray}

\begin{figure}
\includegraphics[width=8.0cm,height=6.0cm]{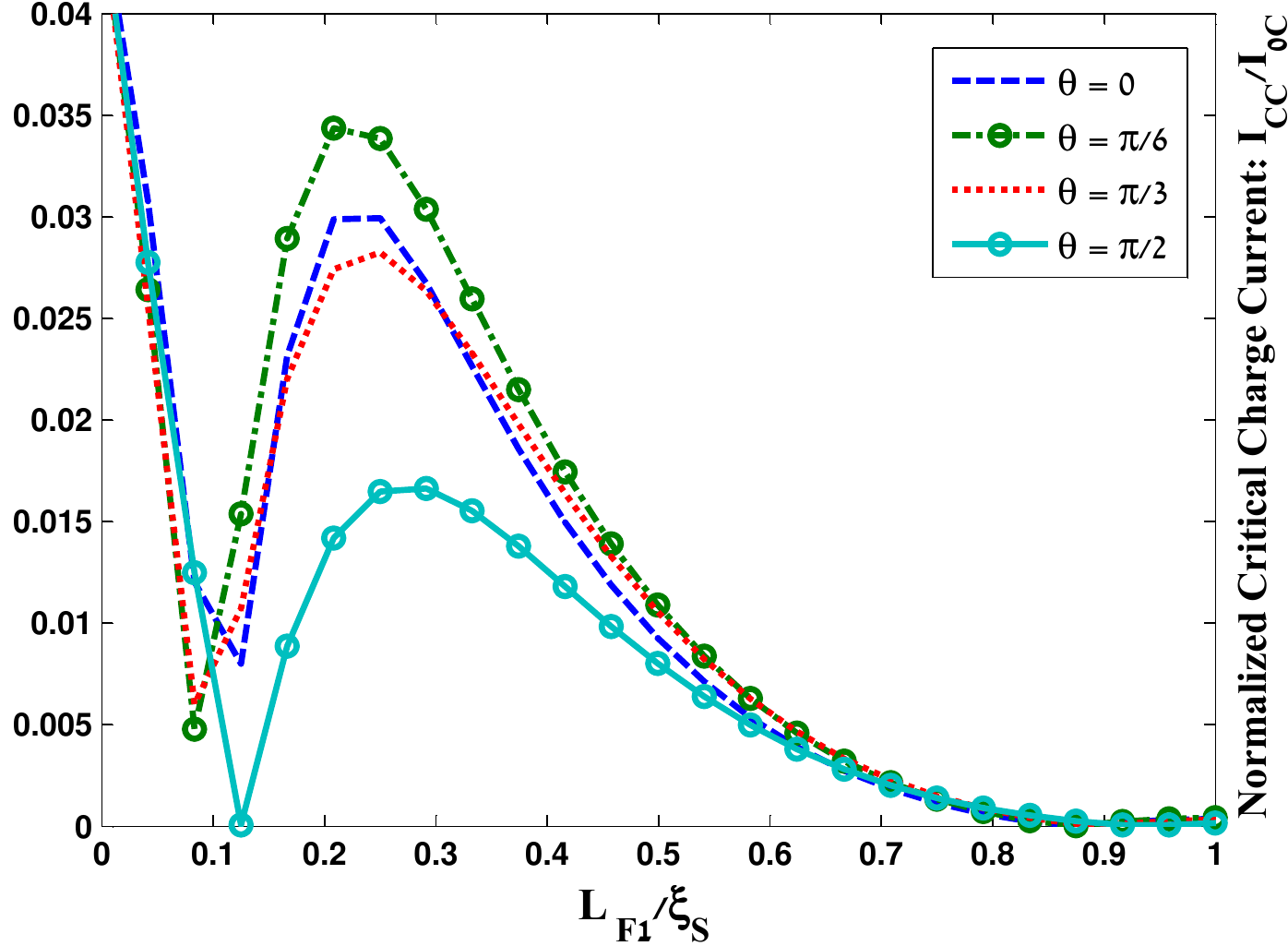}
\caption{\label{fig:Bloch-Co-vsLCo} (Color online) The normalized
critical supercurrent through our proposed
Bloch-wall$\mid$homogeneous ferromagnetic bilayer vs. the length of
homogeneous ferromagnetic layer $L_{\text{F1}}$ for four values of
magnetic orientation angle of homogeneous ferromagnet with respect
to $\hat{z}$-direction, $\theta$=$0$, $\pi/6$, $\pi/3$ and $\pi/2$.
}
\end{figure}
As before, we set the exchange field in the homogeneous ferromagnetic region to $h_1/\Delta_0=15$ whereas it is stronger
in the domain-wall region $h_2/\Delta_0=70$ to avoid influence of
the magnetic field on the domain-wall region. The width of the domain-wall is here set to $L_d = 3.4$ nm, and we show the results for the critical
supercurrent 
in Fig. \ref{fig:Bloch-Co-vsLCo} for four values of magnetic field
orientation angle of the F1 layer, $\theta$=$0$, $\pi/6$, $\pi/3$, and
$\pi/2$. It it seen that the current is enhanced from $\theta$=$0$ up to
values approximately near $\theta$=$\pi/6$, suggesting that the triplet contribution to the current is tuned. The reason for this is that a singlet supercurrent would be completely invariant under a rotation of the exchange field since such a current is spinless. By
increasing value of the angle further from $\theta$=$\pi/6$ up to $\pi/2$,
the critical supercurrent is suppressed. The main advantage of this setup compared to \eg the trilayer structure considered in Ref. \onlinecite{buzdin2} is that only one ferromagnetic layer needs to have its magnetization orientation tuned, which is experimentally more feasible than individually controlling the magnetization structure of each individual layer in a trilayer structure.

\section{Summary}\label{summary}

We have investigated the possibility of establishing a long-range
spin-triplet supercurrent through an inhomogeneous ferromagnetic
region consisting of a Ho$\mid$Co$\mid$Ho trilayer sandwiched
between two conventional $s$-wave superconductors. Utilizing a full
numerical solution in the diffusive regime of transport, the
behavior of the supercurrent in several relevant configurations of
the magnetic trilayer has been obtained. We find qualitatively very
good agreement with the recently reported experimental results by
Robinson \etal \cite{robinson1} regarding the behavior of the
supercurrent as a function of the width of the Co-layer. Moreover,
we find a synthesis of 0-$\pi$ oscillations with superimposed rapid
oscillations when varying the width of the Ho-layer. We are not able
to reproduce the anomalous peaks observed experimentally in this
regime, but note that the results obtained are quite sensitive to
the exact magnetization profile in the Ho-layers. We also investigate the supercurrent in a system
where the intrinsically inhomogeneous Ho ferromagnets are replaced
with domain-wall ferromagnets, and find similar behavior as in the
Ho$\mid$Co$\mid$Ho setup. In addition, we propose a novel type of
ferromagnetic Josephson junction involving a domain-wall and
homogeneous ferromagnet which could be used to obtain a controllable
spin-triplet supercurrent. The advantageous of our proposed
structure compared to the structure considered in Ref.
\onlinecite{buzdin2} is a simpler magnetic profile which could be beneficial from an experimental
point of view.

\section*{Acknowledgement}
We thank J. W. A. Robinson for very useful discussions, and appreciate support from the Physics department computer center of
Isfahan University.

\appendix
\section{\label{pauli}Pauli matrixes}
The Pauli matrices we use in this paper
\begin{align}
\underline{\tau_1} &= \begin{pmatrix}
0 & 1\\
1 & 0\\
\end{pmatrix},\;
\underline{\tau_2} = \begin{pmatrix}
0 & -\i\\
\i & 0\\
\end{pmatrix},\;
\underline{\tau_3} = \begin{pmatrix}
1& 0\\
0& -1\\
\end{pmatrix},\notag\\
\underline{1} &= \begin{pmatrix}
1 & 0\\
0 & 1\\
\end{pmatrix},\;
\hat{1} = \begin{pmatrix}
\underline{1} & \underline{0} \\
\underline{0} & \underline{1} \\
\end{pmatrix},\;
\hat{\tau}_i = \begin{pmatrix}
\underline{\tau_i} & \underline{0}\\
\underline{0} & \underline{\tau_i} \\
\end{pmatrix},\notag\\
\hat{\rho}_1 &= \begin{pmatrix}
\underline{0} & \underline{\tau_1}\\
\underline{\tau_1} & \underline{0} \\
\end{pmatrix},\;
\hat{\rho}_2 =  \begin{pmatrix}
\underline{0} & -i\underline{\tau_1}\\
i\underline{\tau_1} & \underline{0} \\
\end{pmatrix},\;
\hat{\rho}_3 = \begin{pmatrix}
\underline{1} & \underline{0}\\
\underline{0} & -\underline{1}  \\
\end{pmatrix}.
\end{align}

\end{document}